\documentclass[aps,prl,reprint,showpacs,groupedaddress]{revtex4-1}

\usepackage[utf8]{inputenc}
\usepackage[T1]{fontenc}

\usepackage{times}

\usepackage{amsmath}
\usepackage{amsfonts}
\usepackage{amssymb}

\usepackage[pdftex]{graphicx}

\usepackage{hyperref}
\hypersetup{pdfauthor={E. Nicklas, H. Strobel, T. Zibold, C. Gross, B. A. Malomed, P. G. Kevrekidis, and M. K. Oberthaler},
            pdftitle={Rabi flopping induces spatial demixing dynamics}}

\newcommand{\rb}{\ensuremath{^{87}\text{Rb}}}

\newcommand{\ket}[1]{\ensuremath{| #1 \rangle}}

\begin{document}

\title{Rabi flopping induces spatial demixing dynamics}

\author{E. Nicklas}
\email[]{mixing@matterwave.de}
\author{H. Strobel}
\author{T. Zibold}
\author{C. Gross}
\altaffiliation[Present address: ]{Max-Planck-Institut für Quantenoptik, Hans-Kopfermann-Straße 1, 85748 Garching, Germany}
\author{B. A. Malomed}
\altaffiliation[Permanent Address: ]{Department of Physical Electronics, Faculty of Engineering, Tel Aviv University, Tel Aviv 69978, Israel}
\author{P. G. Kevrekidis}
\altaffiliation[Permanent Address: ]{Department of Mathematics and Statistics, University of Massachusetts, Amherst Massachusetts 01003-4515, USA}
\author{M. K. Oberthaler}
\affiliation{Kirchhoff Institute for Physics, University of Heidelberg, INF 227, 69120 Heidelberg, Germany}

\bibliographystyle{apsrev4-1}

\date{\today}

\begin{abstract}
We experimentally investigate the mixing/demixing dynamics of Bose-Einstein condensates in the presence of a linear coupling between two internal states. The observed amplitude reduction of the Rabi oscillations can be understood as a result of demixing dynamics of dressed states as experimentally confirmed by reconstructing the spatial profile of dressed state amplitudes. The observations are in quantitative agreement with numerical integration of coupled Gross-Pitaevskii equations without free parameters, which also reveals the criticality of the dynamics on the symmetry of the system. Our observations demonstrate new possibilities for changing effective atomic interactions and studying critical phenomena.
\end{abstract}


\maketitle

Critical phenomena appear in many areas of physics including phase transitions \cite{stanley71} and nonlinear dynamical systems \cite{guckenheimer83}. Their experimental study requires a high level of control in order to quantitatively compare with theoretical predictions.

Multi-component Bose gases featuring miscibility-immiscibility transitions are prototypical systems for the investigation of critical phenomena due to unprecedented experimental control of the relevant parameters. Early experiments with Bose-Einstein condensates revealed demixing as well as mixing dynamics of two \cite{hall98-1} and three-component \cite{stenger98} quantum fluids. In the latter, even spontaneous symmetry breaking and the corresponding pattern formation has been observed \cite{sadler06, kronjaeger10}. While these experiments have been performed with fixed interaction between the components, atomic systems also allow for the control of the interspecies interaction strength via a Feshbach resonance. This has enabled experiments, that clearly demonstrate miscibility-immiscibility transitions \cite{papp08} and study the two-component dynamics in detail \cite{mertes07, anderson09, tojo10}. An alternative approach for the control of interaction properties and the corresponding dynamics in one-dimensional systems has been demonstrated using state-selective transversal confinement \cite{wicke10}. Recently it has been shown, that the miscibility characteristics of spinor gases can be changed using Raman coupling \cite{lin11}.

\begin{figure}[bt]
    \centering
    \includegraphics[width=86mm]{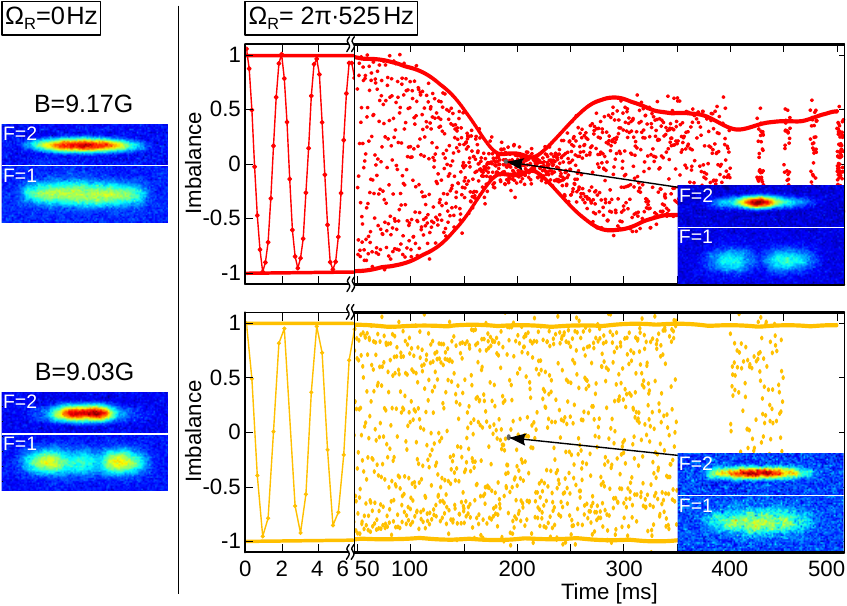}
    \caption{(color online) Rabi oscillations between two hyperfine states of rubidium atoms in the miscible (upper row) and immiscible (lower row) regime of the atomic states. Counter-intuitively the spatially averaged oscillation amplitude is reduced in the miscible regime while it remains close to unity for several hundred cycles in the immiscible case. The false color images contrast the corresponding density distributions of the two components with/without linear coupling indicating that the reduction in amplitude is due to a spatially inhomogeneous phase of the oscillations leading to a variation in the atomic densities at the given time. The different transversal extension of the atomic clouds results from state selective imaging leading to different time-of-flights of the two components. The lines represent the corresponding amplitudes of numerically simulated Rabi oscillations without free parameters.}
    \label{fig_rabi}
\end{figure}

In the present letter, we experimentally investigate the theoretically predicted miscibility properties of two spin states in a Bose-Einstein condensate in the presence of linear coupling \cite{search01, jenkins03, merhasin05}. We report on the experimental observation of the demixing dynamics of the relevant spin states, i.e. dressed states. The (im)miscibility of the system manifests itself in the amplitude of the Rabi oscillations, which is given by the spatial overlap of the corresponding dressed states. Employing both sides of an interspecies Feshbach resonance, the miscible and immiscible regime of the uncoupled two-component system is accessible allowing to contrast the mixing/demixing dynamics to the coupled situation. As shown in the right panel of Fig.\,\ref{fig_rabi} the amplitude of the Rabi oscillations drops in the miscible regime (top) and remains close to unity for immiscible parameters (bottom). These observations indicate a reversal of the miscibility in the presence of a strong linear coupling field.

Before we go into the quantitative discussion of our observations, we provide more details about our experimental system. We prepare a Bose-Einstein condensate of about 4400 \rb{} atoms in the hyperfine state $\ket{1} = \ket{F=1, m_F=1}$ of the ground state manifold confined in an optical dipole trap. Resonant two-photon combined radio-frequency and microwave radiation coherently couples the two hyperfine states $\ket{1}$ and $\ket{2} = \ket{2, -1}$ with a Rabi frequency $\Omega_R = 2 \pi \cdot 525$\,Hz at a detuning of $2 \pi \cdot 200$\,kHz below the intermediate \ket{2, 0} level. The respective intra- and interspecies s-wave scattering lengths of \ket{1} and \ket{2} in units of the Bohr radius are $(a_{11},a_{22},a_{12}) = (95.0, 100.4, 97.7)\,a_{\text{B}}$ \cite{mertes07} leading to a system close to the miscibility-immiscibiliy threshold $a_{12}^2 = a_{11}a_{22}$ \cite{mineev74, timmermans98}.  Utilizing a Feshbach resonance at $B=9.10$\,G \cite{erhard04, widera04} we tune $a_{12}$ into the miscible ($B=9.17$\,G, $a_{12} \approx 94 a_{\text{B}}$) and immiscible regime ($B=9.03$\,G, $a_{12} \approx 102 a_{\text{B}}$) \cite{tojo10}. Three-body recombination and spin relaxation losses in \ket{2} result in a $1/e$-lifetime of $310$\,ms for both magnetic field settings. The quasi one-dimensional confinement with trapping frequencies $(\omega_x, \omega_{\perp}) = 2 \pi \cdot (22, 460)$\,Hz allows for spatial demixing dynamics only along the longitudinal trap axis. The transverse degrees of freedom are frozen because the spin healing length in the trap center $\xi_s \approx 1.2 \,\mu$m is comparable to the transversal extension of the wavefunction of approximately $1.1\,\mu$m. Consecutive absorption imaging (delay of $780\,\mu$s) with high spatial resolution ($1.1\,\mu$m) allows for observation of the atomic density in both hyperfine states for each experimental realization.

\begin{figure}[tb]
    \centering
    \includegraphics[width=86mm]{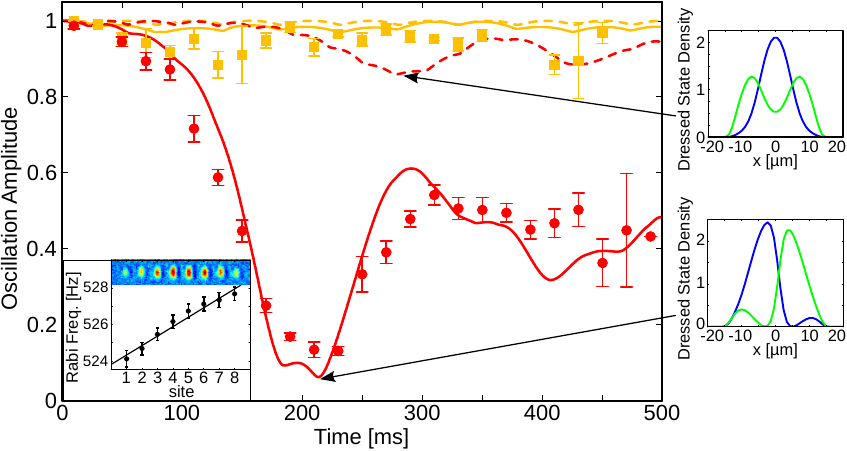}
    \caption{(color online) Quantitative comparison of numerical simulations with experimental observations in the immiscible (orange squares) and miscible (red circles) regimes of the atomic states revealing the criticality of the dynamics. The dashed lines display the result of numerical computations of the full Rabi dynamics. The corresponding dressed state density profiles reveal symmetric demixing as displayed in the upper right inset. The experimentally observed amplitude (solid circles and squares) is only captured if the linear gradient of the Rabi frequency ($2 \pi \cdot 0.94$\,Hz$/10\,\mu$m) is taken into account (solid lines), indicating the criticality of the phenomenon. The gradient has been independently characterized through a local measurement of the Rabi frequency in a lattice as depicted in the left inset. The associated symmetry breaking can be seen in the density distribution of the dressed states (lower right inset).}
    \label{fig_simulations}
\end{figure}

For a quantitative analysis of the observed amplitude characteristics of the Rabi oscillations we theoretically model the dynamics of the wavefunctions of our experimental system, $\psi_1$ and $\psi_2$, by two coupled Gross-Pitaevskii equations including the linear coupling and atom number loss \cite{pitaevskii03, merhasin05, mertes07}.
\begin{align}
    \label{eq_gpe}
    i \hbar \frac{\partial}{\partial t} \psi_1
    & = \left[ -\frac{\hbar^2}{2m} \nabla^2 + V + g_{11} |\psi_1|^2 + g_{12} |\psi_2|^2 \right. \notag \\
    & \qquad{} \left. - i \frac{\hbar}{2} \left( \Gamma_{11} |\psi_1|^2 + \Gamma_{12} |\psi_2|^2 \right) \right] \psi_1 - \frac{\hbar \Omega_R}{2} \psi_2 \notag \\
    i \hbar \frac{\partial}{\partial t} \psi_2
    & = \left[ -\frac{\hbar^2}{2m} \nabla^2 + V + g_{22} |\psi_2|^2 + g_{12} |\psi_1|^2 \right. \notag \\
    & \qquad{} \left. - i \frac{\hbar}{2} \left(\Gamma_{22} |\psi_2|^2 + \Gamma_{12} |\psi_1|^2 \right) \right] \psi_2 - \frac{\hbar \Omega_R}{2} \psi_1
\end{align}
where $2 \pi \hbar$ is Planck's constant, $V$ the external potential, $m$ the atomic mass and $\Omega_R$ the Rabi frequency. For the interaction parameters $g_{ij} = \frac{4 \pi \hbar^2 a_{ij}}{m}$ and loss coefficients $\Gamma_{ij}$ of \rb{} we use the values given in \cite{mertes07}.
Since our confinement is quasi one-dimensional, we employ the nonpolynomial Schrödinger equation (NPSE) \cite{salasnich06} instead of the full three-dimensional description. We numerically integrate the NPSE without free parameters to simulate the full Rabi oscillation dynamics. In Fig.\,\ref{fig_simulations} we compare the predicted amplitude with the experiment, where we deduce this value by extracting the maximum observed amplitude in a time window of one Rabi period and averaging over ten such cycles. The error bars correspond to 1-s.d. statistical uncertainties of the mean value. While we find very good agreement in the immiscible regime of the atomic states with the experimental data, these simulations do not correctly capture the observed amplitude reduction in the miscible regime.

We attribute the deviation of the NPSE predictions from the experimental data in the miscible regime to a spatial dependence of the linear coupling strength $\Omega_R(x)$. We experimentally probe this by adding an optical standing wave potential with a lattice period of $5.5\,\mu$m, which splits the elongated condensate into eight independent lattice sites. Measuring the local resonant Rabi frequency in each site we find a gradient $\kappa \equiv \nabla \Omega_R(x) \approx 2 \pi \cdot 0.94$\,Hz/$10\,\mu$m along the longitudinal trap axis due to a slightly inhomogeneous radio-frequency field (inset of Fig.\,\ref{fig_simulations}). Using Ramsey spectroscopy in the lattice sites, we independently checked that this spatial variation in $\Omega_R(x)$ does not result from a local detuning, e.g. due to magnetic field gradients, which amounts for $\nabla \Omega^{\text{eff}}_{R}(x) < 2 \pi \cdot 0.001$\,Hz/$10\,\mu$m. When including this gradient $\kappa$ in the simulations, very good agreement with our experimental data is found (solid lines in Fig.\,\ref{fig_rabi} and  Fig.\,\ref{fig_simulations}).

In order to provide an intuitive explanation for our observations, we introduce dressed states $\ket{+} = \frac{1}{\sqrt{2}} (\ket{1} + \ket{2})$ and $\ket{-} = \frac{1}{\sqrt{2}} (\ket{1} - \ket{2})$. These states are eigenstates of the linear coupling Hamiltonian with eigenenergies $\pm \frac{\hbar}{2} \Omega_R$. In this picture, resonant Rabi oscillations between atomic states are the result of an interference between equally populated dressed states. The amplitude of the Rabi oscillations is thus given by the spatial overlap of the dressed states \cite{jenkins03}.

The spatial dynamics of the dressed states results from their effective interaction as can be seen by rewriting the equations of motion (Eq.\,\ref{eq_gpe}) in this basis. With that, the linear coupling terms vanish and the atomic scattering lengths $a_{11}, a_{22}, a_{12}$ are replaced by effective dressed state scattering lengths $a_{++} = a_{--} = \frac{1}{4}(a_{11} + a_{22} + 2a_{12})$ and $a_{+-} = \frac{1}{2}(a_{11} + a_{22})$ \cite{search01, jenkins03}. Thus, the slow dynamics of the Rabi oscillation amplitude (see Fig.\ref{fig_rabi}) can be understood as mixing/demixing dynamics of the dressed states. The condition for their stability against demixing reads $a_{+-}^2 < a_{++} a_{--}$ , which in the bare state basis corresponds to $a_{12} > \frac{1}{2}(a_{11} + a_{22})$. Thus, for equal intraspecies scattering lengths $a_{11} = a_{22}$, which is a good approximation for \rb{}, the miscibility conditions for bare and dressed states are mutually exclusive - dressed states are immiscible where bare states are miscible and vice versa.

Numerical simulations provide access to the bare state wavefunctions allowing for direct calculation of the spatial dressed state profiles. In the immiscible regime of the dressed states, ignoring the gradient in the coupling strength leads to spatially symmetric component separation (top right inset of Fig.\,\ref{fig_simulations}). However, the gradient in the linear coupling strength breaks the symmetry leading to biased antisymmetric demixing (bottom right inset of Fig.\,\ref{fig_simulations}). This can be understood as a result of an additional linear potential with opposite slopes for the two dressed states, $V_{\pm} = V \pm \frac{\hbar}{2} \kappa x$, resulting in an equal but opposite shift of their effective potential minima by $\pm 11$\,nm. The qualitative change in the demixing dynamics of the dressed states in response to a small perturbation of the unbiased symmetric configuration demonstrates the criticality of this phenomenon. In contrast, in the miscible regime of the dressed states, the effect of the gradient in coupling strength is small as demonstrated by the persistent spatial overlap of the dressed states during the time evolution.

In order to reconstruct the density profiles of the dressed states from the experimental data we analyze the Rabi oscillations at t=190\,ms spatially resolved.  Sinusoidally fitting the local Rabi oscillations yields their local amplitude $A(x)$ and phase $\phi(x)$. The relative phase of the dressed states is directly given by the fitted phase $\phi(x)$ of the Rabi oscillations. Their amplitude profiles can be inferred using $A(x) = |\sin(2 \alpha(x))|$, with $\alpha$ being the local mixing angle of a superposition of dressed states $\cos{\alpha} \ket{+} + \sin{\alpha} \exp(i \Omega_R t) \ket{-}$. Due to the $\pi/2$ periodicity of $A$ in $\alpha$ it is not possible to unambiguously assign the calculated amplitudes to the dressed states. However, using the fact that a phase jump of $\pi$ corresponds to a node in the amplitude of one of the dressed states and assuming approximately equal populations of the two components, the probability amplitude profiles of the dressed states can be reconstructed.

\begin{figure}[tb]
    \centering
    \includegraphics[width=86mm]{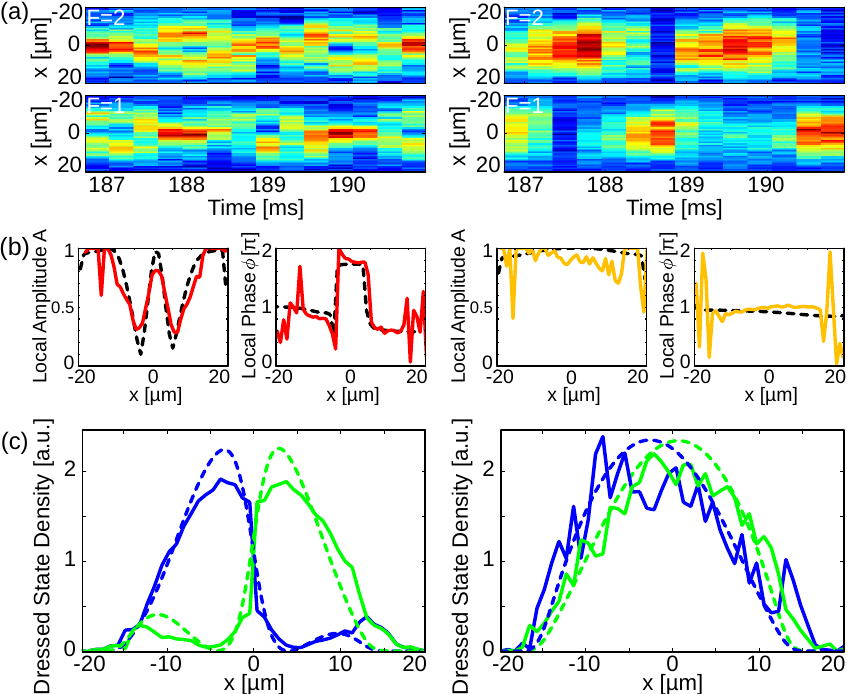}
    \caption{(color online) Reconstruction of dressed state profiles at $B=9.17$\,G (left panel) and $B=9.03$\,G (right). (a) Spatially resolved Rabi oscillations rendered in false color around $t \approx 190$\,ms. In the dressed state immiscible regime (left panel) the oscillations in the center of the cloud are out of phase with respect to its edges, while no spatial dependence is found at miscible case (right panel). (b) A sinusoidal fit to the local Rabi oscillations allows for the determination of their amplitude and phase  (solid red and orange lines). (c) From these measurements we infer the density profiles of the two dressed states (solid blue and green lines). The results of the numerical calculations are shown as dashed lines in (b) and (c).}
    \label{fig_dressed_amplitudes}
\end{figure}

\begin{figure}[tb]
    \centering
    \includegraphics[width=86mm]{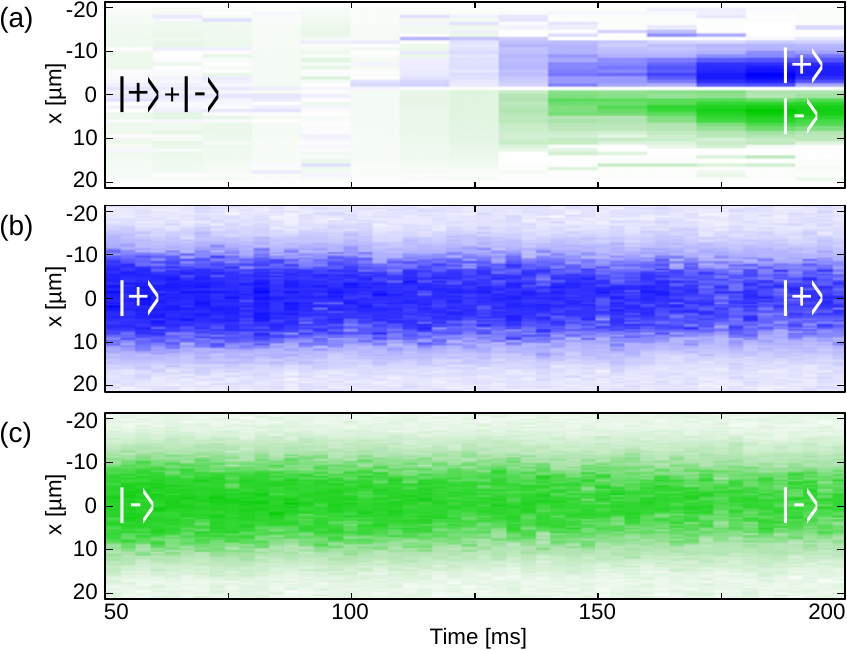}
    \caption{(color online) Time evolution of a superposition of dressed states and of single dressed states in the immiscible regime. (a) The difference of the dressed state densities reconstructed from the Rabi oscillations using the method outlined in Fig.\,\ref{fig_dressed_amplitudes} reveals the demixing dynamics of an initially overlapping superposition of dressed states due to their effective interactions. (b, c) The time evolutions of  initially prepared \ket{+} and \ket{-} states are shown confirming that they are stationary under the action of the linear coupling Hamiltonian.}
    \label{fig_dressed_timetrace}
\end{figure}

The result of the reconstruction for the immiscible regime of dressed states ($B=9.17$\,G) is shown in the left panel of Fig.\,\ref{fig_dressed_amplitudes}. The edge of the atomic cloud oscillates out of phase with the center as can be seen in Fig.\,\ref{fig_dressed_amplitudes}(a) resulting from phase separation of dressed states (Fig.\,\ref{fig_dressed_amplitudes}(c)). On the contrary, in the miscible regime of dressed states $(B=9.03$\,G) neither amplitude nor phase of the Rabi oscillations vary in space (right panel of Fig.\,\ref{fig_dressed_amplitudes}). There, the spatial overlap of the inferred dressed state profiles is only slightly decreased by the gradient in coupling strength demonstrating the miscibility of the dressed states and confirming that the gradient in the linear coupling strength is only a small perturbation. The difference of $\approx 4\,\mu$m in the maxima of the dressed state densities is increased compared to the shift of the effective potentials due to the remaining repulsive interactions between the miscible dressed states. Due to the lower atom density at the edge of the atomic cloud the fit results show increased noise. Using the procedure outlined above, we reconstruct the temporally resolved demixing dynamics shown in Fig.\,\ref{fig_dressed_timetrace}(a).

In order to analyze the stability of single dressed states, we investigate their time evolution. Our system allows their preparation by applying a $\pi/2$ coupling pulse creating an equal superposition of atomic states, followed by a non-adiabatic phase shift of the linear coupling field by $+\frac{\pi}{2}$ ($-\frac{\pi}{2}$) corresponding to the generation of \ket{+} (\ket{-}) dressed states. Note, that in the context of weakly coupled two-mode systems this corresponds to zero-amplitude plasma (pi) oscillations \cite{zibold10}. During the following time evolution we observe that the overlap of the atomic states remains close to unity. Since no Rabi oscillations are observed, an independent measurement of the phase is necessary. This is achieved by employing an additional $\pi/2$ coupling pulse just before readout revealing a homogeneous and temporally constant relative phase between the atomic states. Combining these observations we confirm the stability of single dressed states (Fig.\,\ref{fig_dressed_timetrace}(b, c)). The same observations are made when performing the experiment at $B=9.03$\,G where atomic states demix in the absence of driving. Thus, we experimentally confirm that linear coupling stabilizes an immiscible superposition of atomic states as predicted in \cite{merhasin05}.

To conclude, we have experimentally investigated the miscibility properties of dressed states by determining their density profiles both in the miscible and immiscible regime. The experimental observations are in very good agreement with numerical simulations without free parameters. Comparison with theoretical predictions reveals the criticality of the demixing dynamics on the symmetry of the system. In addition, we have experimentally confirmed that linear coupling stabilizes immiscible two-component gases. In this system one can realize equal interspecies interactions in the strong driving limit, allowing for the experimental exploration of analytically solvable problems, for example in the context of 1D two-component Bose gases \cite{fuchs05}. Preliminary analysis indicates that the intermediate regime of weak linear coupling, where neither the bare nor the dressed states form an appropriate basis, presents a wealth of unexplored nonlinear states with a delicate bifurcation structure. In the context of phase transitions, the demonstrated demixing control has a direct application for tests of the Kibble-Zurek mechanism leading to topological defect formation as proposed in \cite{sabbatini11}. The suggested experiment requires a  miscible-immiscible transition with a gapped energy spectrum which is not available with standard Feshbach tuning.

\begin{acknowledgments}
We gratefully thank Wolfgang M. Müssel for his support during the final steps of the measurements and David Hume for the careful reading of the manuscript. We acknowledge support from the Heidelberg Center for Quantum Dynamics, the DFG-Forschergruppe 760, the German-Israeli Foundation, the ExtreMe Matter Institute, and EU FET-Open project MIDAS. T. Z. acknowledges support from the Landesgraduiertenförderung Baden-Württemberg.
\end{acknowledgments}


\begin{thebibliography}{10}%
\makeatletter
\providecommand \@ifxundefined [1]{%
 \ifx #1\undefined \expandafter \@firstoftwo
 \else \expandafter \@secondoftwo
\fi
}%
\providecommand \@ifnum [1]{%
 \ifnum #1\expandafter \@firstoftwo
 \else \expandafter \@secondoftwo
\fi
}%
\providecommand \enquote [1]{``#1''}%
\providecommand \bibnamefont  [1]{#1}%
\providecommand \bibfnamefont [1]{#1}%
\providecommand \citenamefont [1]{#1}%
\providecommand\href[0]{\@sanitize\@href}%
\providecommand\@href[1]{\endgroup\@@startlink{#1}\endgroup\@@href}%
\providecommand\@@href[1]{#1\@@endlink}%
\providecommand \@sanitize [0]{\begingroup\catcode`\&12\catcode`\#12\relax}%
\@ifxundefined \pdfoutput {\@firstoftwo}{%
 \@ifnum{\z@=\pdfoutput}{\@firstoftwo}{\@secondoftwo}%
}{%
 \providecommand\@@startlink[1]{\leavevmode}%
 \providecommand\@@endlink[0]{}%
}{%
 \providecommand\@@startlink[1]{%
  \leavevmode
  \pdfstartlink
   attr{/Border[0 0 1 ]/H/I/C[0 1 1]}%
   user{/Subtype/Link/A<</Type/Action/S/URI/URI(#1)>>}%
  \relax
 }%
 \providecommand\@@endlink[0]{\pdfendlink}%
}%
\providecommand \url  [0]{\begingroup\@sanitize \@url }%
\providecommand \@url [1]{\endgroup\@href {#1}{\urlprefix}}%
\providecommand \urlprefix [0]{URL }%
\providecommand \Eprint[0]{\href }%
\@ifxundefined \urlstyle {%
  \providecommand \doi [1]{doi:\discretionary{}{}{}#1}%
}{%
  \providecommand \doi [0]{doi:\discretionary{}{}{}\begingroup
  \urlstyle{rm}\Url }%
}%
\providecommand \doibase [0]{http://dx.doi.org/}%
\providecommand \Doi[1]{\href{\doibase#1}}%
\providecommand \bibAnnote [3]{%
  \BibitemShut{#1}%
  \begin{quotation}\noindent
    \textsc{Key:}\ #2\\\textsc{Annotation:}\ #3%
  \end{quotation}%
}%
\providecommand \bibAnnoteFile [2]{%
  \IfFileExists{#2}{\bibAnnote {#1} {#2} {\input{#2}}}{}%
}%
\providecommand \typeout [0]{\immediate \write \m@ne }%
\providecommand \selectlanguage [0]{\@gobble}%
\providecommand \bibinfo [0]{\@secondoftwo}%
\providecommand \bibfield [0]{\@secondoftwo}%
\providecommand \translation [1]{[#1]}%
\providecommand \BibitemOpen[0]{}%
\providecommand \bibitemStop [0]{}%
\providecommand \bibitemNoStop [0]{.\EOS\space}%
\providecommand \EOS [0]{\spacefactor3000\relax}%
\providecommand \BibitemShut [1]{\csname bibitem#1\endcsname}%
\bibitem{stanley71}%
  \BibitemOpen
  \bibfield{author}{%
  \bibinfo {author} {\bibfnamefont{H.~E.}\ \bibnamefont{Stanley}},\ }%
  \emph{\bibinfo {title} {{Introduction to phase transitions and critical
  phenomena}}}\ (\bibinfo {publisher} {{Oxford University Press}},\ \bibinfo
  {year} {{1971}})%
  \bibAnnoteFile{NoStop}{stanley71}%
\bibitem{guckenheimer83}%
  \BibitemOpen
  \bibfield{author}{%
  \bibinfo {author} {\bibfnamefont{J.}~\bibnamefont{Guckenheimer}}\ and\
  \bibinfo {author} {\bibfnamefont{P.}~\bibnamefont{Holmes}},\ }%
  \emph{\bibinfo {title} {{Nonlinear Oscillations, Dynamical Systems, and
  Bifurcations of Vector Fields}}}\ (\bibinfo {publisher} {{Springer}},\
  \bibinfo {year} {{1983}})%
  \bibAnnoteFile{NoStop}{guckenheimer83}%
\bibitem{hall98-1}%
  \BibitemOpen
  \bibfield{author}{%
  \bibinfo {author} {\bibfnamefont{D.~S.}\ \bibnamefont{Hall}}, \bibinfo
  {author} {\bibfnamefont{M.~R.}\ \bibnamefont{Matthews}}, \bibinfo {author}
  {\bibfnamefont{J.~R.}\ \bibnamefont{Ensher}}, \bibinfo {author}
  {\bibfnamefont{C.~E.}\ \bibnamefont{Wieman}},\ and\ \bibinfo {author}
  {\bibfnamefont{E.~A.}\ \bibnamefont{Cornell}},\ }%
  \bibfield{journal}{%
  \Doi{{10.1103/PhysRevLett.81.1539}}{\bibinfo {journal} {{Phys. Rev. Lett.}}}\
  }%
  \textbf{\bibinfo {volume} {{81}}},\ \bibinfo {pages} {1539} (\bibinfo {year}
  {{1998}})%
  \bibAnnoteFile{NoStop}{hall98-1}%
\bibitem{stenger98}%
  \BibitemOpen
  \bibfield{author}{%
  \bibinfo {author} {\bibfnamefont{J.}~\bibnamefont{Stenger}}, \bibinfo
  {author} {\bibfnamefont{S.}~\bibnamefont{Inouye}}, \bibinfo {author}
  {\bibfnamefont{D.~M.}\ \bibnamefont{Stamper-Kurn}}, \bibinfo {author}
  {\bibfnamefont{H.~J.}\ \bibnamefont{Miesner}}, \bibinfo {author}
  {\bibfnamefont{A.~P.}\ \bibnamefont{Chikkatur}},\ and\ \bibinfo {author}
  {\bibfnamefont{W.}~\bibnamefont{Ketterle}},\ }%
  \bibfield{journal}{%
  \Doi{{10.1038/24567}}{\bibinfo {journal} {{Nature}}}\ }%
  \textbf{\bibinfo {volume} {{396}}},\ \bibinfo {pages} {345} (\bibinfo {year}
  {{1998}})%
  \bibAnnoteFile{NoStop}{stenger98}%
\bibitem{sadler06}%
  \BibitemOpen
  \bibfield{author}{%
  \bibinfo {author} {\bibfnamefont{L.~E.}\ \bibnamefont{Sadler}}, \bibinfo
  {author} {\bibfnamefont{J.~M.}\ \bibnamefont{Higbie}}, \bibinfo {author}
  {\bibfnamefont{S.~R.}\ \bibnamefont{Leslie}}, \bibinfo {author}
  {\bibfnamefont{M.}~\bibnamefont{Vengalattore}},\ and\ \bibinfo {author}
  {\bibfnamefont{D.~M.}\ \bibnamefont{Stamper-Kurn}},\ }%
  \bibfield{journal}{%
  \Doi{{10.1038/nature05094}}{\bibinfo {journal} {{Nature}}}\ }%
  \textbf{\bibinfo {volume} {{443}}},\ \bibinfo {pages} {312} (\bibinfo {year}
  {{2006}})%
  \bibAnnoteFile{NoStop}{sadler06}%
\bibitem{kronjaeger10}%
  \BibitemOpen
  \bibfield{author}{%
  \bibinfo {author} {\bibfnamefont{J.}~\bibnamefont{Kronj\"ager}}, \bibinfo
  {author} {\bibfnamefont{C.}~\bibnamefont{Becker}}, \bibinfo {author}
  {\bibfnamefont{P.}~\bibnamefont{Soltan-Panahi}}, \bibinfo {author}
  {\bibfnamefont{K.}~\bibnamefont{Bongs}},\ and\ \bibinfo {author}
  {\bibfnamefont{K.}~\bibnamefont{Sengstock}},\ }%
  \bibfield{journal}{%
  \Doi{{10.1103/PhysRevLett.105.090402}}{\bibinfo {journal} {{Phys. Rev.
  Lett.}}}\ }%
  \textbf{\bibinfo {volume} {{105}}},\ \bibinfo {pages} {090402} (\bibinfo
  {year} {{2010}})%
  \bibAnnoteFile{NoStop}{kronjaeger10}%
\bibitem{papp08}%
  \BibitemOpen
  \bibfield{author}{%
  \bibinfo {author} {\bibfnamefont{S.~B.}\ \bibnamefont{Papp}}, \bibinfo
  {author} {\bibfnamefont{J.~M.}\ \bibnamefont{Pino}},\ and\ \bibinfo {author}
  {\bibfnamefont{C.~E.}\ \bibnamefont{Wieman}},\ }%
  \bibfield{journal}{%
  \Doi{{10.1103/PhysRevLett.101.040402}}{\bibinfo {journal} {{Phys. Rev.
  Lett.}}}\ }%
  \textbf{\bibinfo {volume} {{101}}},\ \bibinfo {pages} {040402} (\bibinfo
  {year} {{2008}})%
  \bibAnnoteFile{NoStop}{papp08}%
\bibitem{mertes07}%
  \BibitemOpen
  \bibfield{author}{%
  \bibinfo {author} {\bibfnamefont{K.~M.}\ \bibnamefont{Mertes}}, \bibinfo
  {author} {\bibfnamefont{J.~W.}\ \bibnamefont{Merrill}}, \bibinfo {author}
  {\bibfnamefont{R.}~\bibnamefont{Carretero-Gonz\'alez}}, \bibinfo {author}
  {\bibfnamefont{D.~J.}\ \bibnamefont{Frantzeskakis}}, \bibinfo {author}
  {\bibfnamefont{P.~G.}\ \bibnamefont{Kevrekidis}},\ and\ \bibinfo {author}
  {\bibfnamefont{D.~S.}\ \bibnamefont{Hall}},\ }%
  \bibfield{journal}{%
  \Doi{{10.1103/PhysRevLett.99.190402}}{\bibinfo {journal} {{Phys. Rev.
  Lett.}}}\ }%
  \textbf{\bibinfo {volume} {{99}}},\ \bibinfo {pages} {190402} (\bibinfo
  {year} {{2007}})%
  \bibAnnoteFile{NoStop}{mertes07}%
\bibitem{anderson09}%
  \BibitemOpen
  \bibfield{author}{%
  \bibinfo {author} {\bibfnamefont{R.~P.}\ \bibnamefont{Anderson}}, \bibinfo
  {author} {\bibfnamefont{C.}~\bibnamefont{Ticknor}}, \bibinfo {author}
  {\bibfnamefont{A.~I.}\ \bibnamefont{Sidorov}},\ and\ \bibinfo {author}
  {\bibfnamefont{B.~V.}\ \bibnamefont{Hall}},\ }%
  \bibfield{journal}{%
  \Doi{{10.1103/PhysRevA.80.023603}}{\bibinfo {journal} {{Phys. Rev. A}}}\ }%
  \textbf{\bibinfo {volume} {{80}}},\ \bibinfo {pages} {023603} (\bibinfo
  {year} {{2009}})%
  \bibAnnoteFile{NoStop}{anderson09}%
\bibitem{tojo10}%
  \BibitemOpen
  \bibfield{author}{%
  \bibinfo {author} {\bibfnamefont{S.}~\bibnamefont{Tojo}}, \bibinfo {author}
  {\bibfnamefont{Y.}~\bibnamefont{Taguchi}}, \bibinfo {author}
  {\bibfnamefont{Y.}~\bibnamefont{Masuyama}}, \bibinfo {author}
  {\bibfnamefont{T.}~\bibnamefont{Hayashi}}, \bibinfo {author}
  {\bibfnamefont{H.}~\bibnamefont{Saito}},\ and\ \bibinfo {author}
  {\bibfnamefont{T.}~\bibnamefont{Hirano}},\ }%
  \bibfield{journal}{%
  \Doi{{10.1103/PhysRevA.82.033609}}{\bibinfo {journal} {{Phys. Rev. A}}}\ }%
  \textbf{\bibinfo {volume} {{82}}},\ \bibinfo {pages} {033609} (\bibinfo
  {year} {{2010}})%
  \bibAnnoteFile{NoStop}{tojo10}%
\bibitem{wicke10}%
  \BibitemOpen
  \bibfield{author}{%
  \bibinfo {author} {\bibfnamefont{P.}~\bibnamefont{Wicke}}, \bibinfo {author}
  {\bibfnamefont{S.}~\bibnamefont{Whitlock}},\ and\ \bibinfo {author}
  {\bibfnamefont{N.~J.~v.}\ \bibnamefont{Druten}}}%
   (\bibinfo {year} {{2010}}),\
  \Eprint{http://arxiv.org/abs/{1010.4545v1}}{{arXiv}:{1010.4545v1}}%
  \bibAnnoteFile{NoStop}{wicke10}%
\bibitem{lin11}%
  \BibitemOpen
  \bibfield{author}{%
  \bibinfo {author} {\bibfnamefont{Y.-J.}\ \bibnamefont{Lin}}, \bibinfo
  {author} {\bibfnamefont{K.}~\bibnamefont{Jiménez-García}},\ and\ \bibinfo
  {author} {\bibfnamefont{I.~B.}\ \bibnamefont{Spielman}},\ }%
  \bibfield{journal}{%
  \Doi{{10.1038/nature09887}}{\bibinfo {journal} {{Nature}}}\ }%
  \textbf{\bibinfo {volume} {{471}}},\ \bibinfo {pages} {83} (\bibinfo {year}
  {{2011}})%
  \bibAnnoteFile{NoStop}{lin11}%
\bibitem{search01}%
  \BibitemOpen
  \bibfield{author}{%
  \bibinfo {author} {\bibfnamefont{C.~P.}\ \bibnamefont{Search}}\ and\ \bibinfo
  {author} {\bibfnamefont{P.~R.}\ \bibnamefont{Berman}},\ }%
  \bibfield{journal}{%
  \Doi{{10.1103/PhysRevA.63.043612}}{\bibinfo {journal} {{Phys. Rev. A}}}\ }%
  \textbf{\bibinfo {volume} {{63}}},\ \bibinfo {pages} {043612} (\bibinfo
  {year} {{2001}})%
  \bibAnnoteFile{NoStop}{search01}%
\bibitem{jenkins03}%
  \BibitemOpen
  \bibfield{author}{%
  \bibinfo {author} {\bibfnamefont{S.~D.}\ \bibnamefont{Jenkins}}\ and\
  \bibinfo {author} {\bibfnamefont{T.~A.~B.}\ \bibnamefont{Kennedy}},\ }%
  \bibfield{journal}{%
  \Doi{{10.1103/PhysRevA.68.053607}}{\bibinfo {journal} {{Phys. Rev. A}}}\ }%
  \textbf{\bibinfo {volume} {{68}}},\ \bibinfo {pages} {053607} (\bibinfo
  {year} {{2003}})%
  \bibAnnoteFile{NoStop}{jenkins03}%
\bibitem{merhasin05}%
  \BibitemOpen
  \bibfield{author}{%
  \bibinfo {author} {\bibfnamefont{I.~M.}\ \bibnamefont{Merhasin}}, \bibinfo
  {author} {\bibfnamefont{B.~A.}\ \bibnamefont{Malomed}},\ and\ \bibinfo
  {author} {\bibfnamefont{R.}~\bibnamefont{Driben}},\ }%
  \bibfield{journal}{%
  \Doi{{10.1088/0953-4075/38/7/009}}{\bibinfo {journal} {{Journal of Physics B:
  Atomic, Molecular and Optical Physics}}}\ }%
  \textbf{\bibinfo {volume} {{38}}},\ \bibinfo {pages} {877} (\bibinfo {year}
  {{2005}})%
  \bibAnnoteFile{NoStop}{merhasin05}%
\bibitem{mineev74}%
  \BibitemOpen
  \bibfield{author}{%
  \bibinfo {author} {\bibfnamefont{V.~P.}\ \bibnamefont{Mineev}},\ }%
  \bibfield{journal}{%
  \bibinfo {journal} {{Sov. Phys. JETP}}\ }%
  \textbf{\bibinfo {volume} {{40}}},\ \bibinfo {pages} {132} (\bibinfo {year}
  {{1974}})%
  \bibAnnoteFile{NoStop}{mineev74}%
\bibitem{timmermans98}%
  \BibitemOpen
  \bibfield{author}{%
  \bibinfo {author} {\bibfnamefont{E.}~\bibnamefont{Timmermans}},\ }%
  \bibfield{journal}{%
  \Doi{{10.1103/PhysRevLett.81.5718}}{\bibinfo {journal} {{Phys. Rev. Lett.}}}\
  }%
  \textbf{\bibinfo {volume} {{81}}},\ \bibinfo {pages} {5718} (\bibinfo {year}
  {{1998}})%
  \bibAnnoteFile{NoStop}{timmermans98}%
\bibitem{erhard04}%
  \BibitemOpen
  \bibfield{author}{%
  \bibinfo {author} {\bibfnamefont{M.}~\bibnamefont{Erhard}}, \bibinfo {author}
  {\bibfnamefont{H.}~\bibnamefont{Schmaljohann}}, \bibinfo {author}
  {\bibfnamefont{J.}~\bibnamefont{Kronj\"ager}}, \bibinfo {author}
  {\bibfnamefont{K.}~\bibnamefont{Bongs}},\ and\ \bibinfo {author}
  {\bibfnamefont{K.}~\bibnamefont{Sengstock}},\ }%
  \bibfield{journal}{%
  \Doi{{10.1103/PhysRevA.69.032705}}{\bibinfo {journal} {{Phys. Rev. A}}}\ }%
  \textbf{\bibinfo {volume} {{69}}},\ \bibinfo {pages} {032705} (\bibinfo
  {year} {{2004}})%
  \bibAnnoteFile{NoStop}{erhard04}%
\bibitem{widera04}%
  \BibitemOpen
  \bibfield{author}{%
  \bibinfo {author} {\bibfnamefont{A.}~\bibnamefont{Widera}}, \bibinfo {author}
  {\bibfnamefont{O.}~\bibnamefont{Mandel}}, \bibinfo {author}
  {\bibfnamefont{M.}~\bibnamefont{Greiner}}, \bibinfo {author}
  {\bibfnamefont{S.}~\bibnamefont{Kreim}}, \bibinfo {author}
  {\bibfnamefont{T.~W.}\ \bibnamefont{H\"ansch}},\ and\ \bibinfo {author}
  {\bibfnamefont{I.}~\bibnamefont{Bloch}},\ }%
  \bibfield{journal}{%
  \Doi{{10.1103/PhysRevLett.92.160406}}{\bibinfo {journal} {{Phys. Rev.
  Lett.}}}\ }%
  \textbf{\bibinfo {volume} {{92}}},\ \bibinfo {pages} {160406} (\bibinfo
  {year} {{2004}})%
  \bibAnnoteFile{NoStop}{widera04}%
\bibitem{pitaevskii03}%
  \BibitemOpen
  \bibfield{author}{%
  \bibinfo {author} {\bibfnamefont{L.}~\bibnamefont{Pitaevskii}}\ and\ \bibinfo
  {author} {\bibfnamefont{S.}~\bibnamefont{Stringari}},\ }%
  \emph{\bibinfo {title} {{Bose-Einstein Condensation}}}\ (\bibinfo {publisher}
  {{Oxford University Press}},\ \bibinfo {year} {{2003}})%
  \bibAnnoteFile{NoStop}{pitaevskii03}%
\bibitem{salasnich06}%
  \BibitemOpen
  \bibfield{author}{%
  \bibinfo {author} {\bibfnamefont{L.}~\bibnamefont{Salasnich}}\ and\ \bibinfo
  {author} {\bibfnamefont{B.~A.}\ \bibnamefont{Malomed}},\ }%
  \bibfield{journal}{%
  \Doi{{10.1103/PhysRevA.74.053610}}{\bibinfo {journal} {{Phys. Rev. A}}}\ }%
  \textbf{\bibinfo {volume} {{74}}},\ \bibinfo {pages} {053610} (\bibinfo
  {year} {{2006}})%
  \bibAnnoteFile{NoStop}{salasnich06}%
\bibitem{zibold10}%
  \BibitemOpen
  \bibfield{author}{%
  \bibinfo {author} {\bibfnamefont{T.}~\bibnamefont{Zibold}}, \bibinfo {author}
  {\bibfnamefont{E.}~\bibnamefont{Nicklas}}, \bibinfo {author}
  {\bibfnamefont{C.}~\bibnamefont{Gross}},\ and\ \bibinfo {author}
  {\bibfnamefont{M.~K.}\ \bibnamefont{Oberthaler}},\ }%
  \bibfield{journal}{%
  \Doi{{10.1103/PhysRevLett.105.204101}}{\bibinfo {journal} {{Phys. Rev.
  Lett.}}}\ }%
  \textbf{\bibinfo {volume} {{105}}},\ \bibinfo {pages} {204101} (\bibinfo
  {year} {{2010}})%
  \bibAnnoteFile{NoStop}{zibold10}%
\bibitem{fuchs05}%
  \BibitemOpen
  \bibfield{author}{%
  \bibinfo {author} {\bibfnamefont{J.~N.}\ \bibnamefont{Fuchs}}, \bibinfo
  {author} {\bibfnamefont{D.~M.}\ \bibnamefont{Gangardt}}, \bibinfo {author}
  {\bibfnamefont{T.}~\bibnamefont{Keilmann}},\ and\ \bibinfo {author}
  {\bibfnamefont{G.~V.}\ \bibnamefont{Shlyapnikov}},\ }%
  \bibfield{journal}{%
  \Doi{{10.1103/PhysRevLett.95.150402}}{\bibinfo {journal} {{Phys. Rev.
  Lett.}}}\ }%
  \textbf{\bibinfo {volume} {{95}}},\ \bibinfo {pages} {150402} (\bibinfo
  {year} {{2005}})%
  \bibAnnoteFile{NoStop}{fuchs05}%
\bibitem{sabbatini11}%
  \BibitemOpen
  \bibfield{author}{%
  \bibinfo {author} {\bibfnamefont{J.}~\bibnamefont{Sabbatini}}, \bibinfo
  {author} {\bibfnamefont{W.~H.}\ \bibnamefont{Zurek}},\ and\ \bibinfo {author}
  {\bibfnamefont{M.~J.}\ \bibnamefont{Davis}}}%
   (\bibinfo {year} {{2011}}),\
  \Eprint{http://arxiv.org/abs/{1106.5843v1}}{{arXiv}:{1106.5843v1}}%
  \bibAnnoteFile{NoStop}{sabbatini11}%
\end{thebibliography}

%

\end{document}